\documentclass[sigconf,10pt]{acmart}

\usepackage{booktabs}
\usepackage{multirow}
\usepackage{wrapfig}

\copyrightyear{2019} 
\acmYear{2019} 
\setcopyright{acmlicensed}
\acmConference[MobiCom '19]{The 25th Annual International Conference on Mobile Computing and Networking}{October 21--25, 2019}{Los Cabos, Mexico}
\acmBooktitle{The 25th Annual International Conference on Mobile Computing and Networking (MobiCom '19), October 21--25, 2019, Los Cabos, Mexico}
\acmPrice{15.00}
\acmDOI{10.1145/3300061.3300132}
\acmISBN{978-1-4503-6169-9/19/10}

\begin{document}
\title{Experience: Understanding Long-Term Evolving\\ Patterns of Shared Electric Vehicle Networks}

\author{Guang Wang}
\titlenote{Prof. D. Zhang and Y. Wang are the joint corresponding authors.}
\affiliation{%
  \institution{Rutgers University}
}
\email{guang.wang@rutgers.edu}

\author{Xiuyuan Chen}
\affiliation{%
  \institution{Rutgers University}
}
\email{xc265@rutgers.edu}

\author{Fan Zhang}
\affiliation{
  \institution{Shenzhen Institutes of Advanced Technology, CAS}
}
\email{zhangfan@siat.ac.cn}

\author{Yang Wang}
\affiliation{
  \institution{University of Science and Technology of China}
  }
\email{angyan@ustc.edu.cn}

\author{Desheng Zhang}
\affiliation{
  \institution{Rutgers University}
}
\email{desheng.zhang@cs.rutgers.edu}

\renewcommand{\shortauthors}{G. Wang et al.}

\begin{abstract}
Due to the ever-growing concerns on the air pollution and energy security, many cities have started to update their taxi fleets with electric ones.
Although environmentally friendly, the rapid promotion of electric taxis raises problems to both taxi drivers and governments, e.g., prolonged waiting/charging time, unbalanced utilization of charging infrastructures and reduced taxi supply due to the long charging time.
In this paper, we make the first effort to understand the long-term evolving patterns through a five-year study on one of the largest electric taxi networks in the world, i.e., the Shenzhen electric taxi network in China.
In particular, we perform a comprehensive measurement investigation called ePat to explore the evolving mobility and charging patterns of electric vehicles. Our ePat is based on 4.8 TB taxi GPS data, 240 GB taxi transaction data, and metadata from 117 charging stations, during an evolving process from 427 electric taxis in 2013 to 13,178 in 2018.
Moreover, ePat also explores the impacts of various contexts and benefits during the evolving process.
Our ePat as a comprehensive investigation of the electric taxi network mobility and charging evolving has the potential to advance the understanding of the evolving patterns of electric taxi networks and pave the way for analyzing future shared autonomous vehicles.
\end{abstract}

%
%



\maketitle

\vspace*{-5pt}

\section{Introduction}
\label{sec:intro}

In the vision of smart cities, vehicle electrification has become inevitable since it contributes to significant emission reduction for better air quality and less energy consumption \cite{bloomberg2010exploring, bCharge, zhang2014exploring, zhang2015urbancps, zhang2015feeder, zhang2013coride}.
 Compared to personal cars, taxis as shared vehicles have high gas consumption and emissions due to their long-time daily operations. For example, replacing a conventional gas taxi with an Electric Taxi (ET) creates an emission impact that is equivalent to replacing eight New York City personal cars with electric vehicles \cite{taxi2013take}. It provides a higher incentive for city governments to replace conventional gas taxis with ETs.
 In the recent decade, many cities around the world have initiated the process of taxi electrification, e.g., New York City, London, Beijing, and Shenzhen \cite{tu2016optimizing}.
 Among these cities, the Chinese city Shenzhen has started the taxi electrification process from 2010 and achieved the largest ET network in the world by 2017 \cite{SZET}.
 In particular, the number of ETs in Shenzhen has increased from 50 in 2010 to 12,518 in 2017 \cite{SZET1}, and it is projected to be over 18,000 in 2020, becoming an ET-only taxi network.

Understanding such long-term evolving patterns of the Shenzhen ET network is essential for predicting and quantifying ET development roadblocks and benefits for Shenzhen and other cities. The understanding of ET networks may be potential for future shared autonomous vehicles \cite{burns2013sustainable, greenblatt2015autonomous}, which can be treated as ETs without human drivers \cite{kang2017autonomous}.

However, the unique characteristics of ETs, e.g., the long charging durations and unexpected waiting time for refueling compared with gas taxis, make their operation and charging patterns very complicated regarding the 5-year evolving. For example,
the average daily operation distance of ETs in Shenzhen is 430 km, which causes frequent charging activities because of the limited battery capacity, e.g., average 3.5 times per day \cite{tian2014understanding}. Besides, due to the limited charging station supply and uneven charging demand, the average charging time combined with the waiting time for a charging activity is about 1.5 hours, which reduces the overall taxi business time by 13.6\%. Moreover, the energy consumption of ETs varies in different contexts, e.g., the lower mileages caused by the air conditioners at low temperatures.
All these factors, along with a city's ET evolving process (e.g., more ETs, more charging stations, better vehicular make/model), make it extremely challenging to understand the long-term evolving patterns of a large-scale ET network.

In this paper, we perform a comprehensive measurement investigation called ePat to explore the evolving patterns of the ET network based on multi-source datasets.
With ePat, we systematically measure the mobility and charging evolving patterns of ETs, combined with the accessory charging network evolving patterns. Finally, an in-depth discussion is made on the potential applications of our measurement investigation for other cities that plan to promote large-scale ET fleets and future shared autonomous vehicles. In particular, the major contributions of this paper include:
\begin{itemize}
  \item To our best knowledge, we conduct the first work to measure and understand the long-term evolving patterns of electric vehicle networks.
   Our data-driven investigation has three key features:
   (i) a long period, i.e., more than five years;
   (ii) a large number of ETs, i.e., 13,178 ETs;
   (iii) a large number of charging stations and points, e.g., 117 charging stations.
   Such a large-scale data-driven investigation enables us to understand the evolving process of the mobility and charging patterns of the ET network, which are difficult to be achieved with a small-scale or short-term investigation.

  \item We present a measurement investigation called ePat for a systematical context-aware measurement study for ET networks.
  In addition to ET data and charging station data, various contextual data are also leveraged for spatiotemporal modeling, including urban partition, weather conditions, etc. We report our measurement results on various metrics related to spatial, temporal, mobility, energy, CO$_2$ emissions, and drivers' incomes. More importantly, we explain possible causalities based on various correlation analyses.

  \item Based on our measurement results, we provide some in-depth discussions for the lessons and insights learned, including ET and charging network evolving patterns, along with how to apply our experiences to other cities for charging station deployment and policy guidance.
  Some of our data-driven insights have been provided to the Shenzhen transportation committee for a better future ET development.
 \end{itemize}

\section{Related Work}
\label{sec:related}
Most investigations of ETs can be classified into two types based on the investigating periods, i.e., long-term investigation (over one year) and short-term investigation.
For the existing research, some works are based on large-scale real-world data (over 1,000 vehicles), while others leverage a small dataset or simulations.
Based on these two factors, we divide the ET research into four categories, as given in Table~\ref{tab:categories}.

\begin{table}[h!]
\small
\caption{Categories of related works}
\centering
\begin{tabular}{ |c|c|c|}
\hline
Electric Vehicles & Small-Scale & Large-Scale\\
\hline
Short-Term & \cite{DongREC, li2015growing, tian2014understanding, tian2016real, sarker2018morp} & \cite{Yan2018EOC, wang2018toward, yan2017catcharger, liu2016optimal, Du2018DCP, wang2015planning} \\
\hline
Long-Term & \cite{tian2017understanding, zou2016large, qi2018investigating, hu2018analyzing} & ePat (Our work)\\
\hline
\end{tabular}
\vspace*{-12pt}
\label{tab:categories}
\end{table}

\begin{table*}[!ht]\centering
\vspace*{0pt}
\caption{Examples of all data sources}
\begin{tabular}{llllll}
\hline
\multirow{2}{*}{GPS} & plate ID & longitude & latitude & time & speed (km/h) \\ \cline{2-6}
 & TIDXXXX & 114.022901 & 22.532104 & 2016-06-16 08:34:43 & 22 \\ \hline
 \hline
 \multirow{2}{*}{Transaction} & plate ID & pickup time & dropoff time & cost (CNY) & travel distance (m) \\ \cline{2-6}
 & TIDXXXX & 2015-09-03 13:47:58 & 2015-09-03 13:57:23 & 22.6 & 6954 \\ \hline
 \hline
 \multirow{2}{*}{Charging Station}& station ID & station name & longitude & latitude & \# of charging points \\ \cline{2-6}
 & 30 & NB0005 & 113.9878608 & 22.55955418 & 40 \\ \hline
 \hline
  \multirow{2}{*}{Urban Partition}& region ID & longitude1 & latitude1  & longitude2 & latitude2 \\ \cline{2-6}
 & 1 &114.31559657&22.78559093&114.311230763&22.78220351 \\ \hline
\end{tabular}
\vspace*{-5pt}
\label{tab:data}
\end{table*}

\subsection{Short-Term Investigation}
\noindent\textbf{Small-scale}:
Numerous studies \cite{tseng2018improving, ardakanian2012realtime, xiong2015optimal, xiong2016optimal, dandl2018comparing, li2018planning, zhu2018joint, yang2017ev, zhang2018optimal, dai2014stochastic, yang2018charging, sun2017performance, zhang2014charging, you2016optimal, cano2018batteries, yang2018predictive, khonji2018challenges} have been carried out for dealing with real problems in the ET domain, e.g., charging station siting and deployment, charging schedule. However, most of the existing works are based on small-scale real-world ET data or utilize conventional gas taxi data to simulate ET scenarios.
\cite{DongREC} develops a real-time scheduling approach for ET fleets based on GPS trajectory records of 550 ETs.
\cite{li2015growing} develops an optimal charging station deployment framework based on 490 ETs' trajectory records.
Compared to these research, our dataset includes the real-time GPS records of more than 13,000 ETs.

\noindent\textbf{Large-scale:}
\cite{Yan2018EOC} presents an opportunistic wireless charger deployment scheme for ET fleets to minimize the total deployment cost and maximize the opportunity of picking up passengers at the chargers based on a large-scale gas taxi trajectory data.
\cite{wang2018toward} develops a scheduling strategy for future city-scale ETs to avoid congestion in the swapping stations by leveraging the data of the existing conventional taxi fleet.
But they are based on short-term gas taxi data, which is difficult to uncover the operating and charging characteristics of ETs, whereas we study a five-year real-world dataset from a large-scale ET network.

\subsection{Long-Term Investigation}

\noindent\textbf{Small-scale:} There are also some existing works for the long-term investigation of the ET mobility patterns.
\cite{tian2017understanding} investigates four years' GPS trace data from 850 ET in Shenzhen to understand the battery degradation of ETs.
\cite{zou2016large} analyzes a two-year dataset from 34 ETs in Beijing to understand the operational status, benefits, and charging facilities.
These small networks cannot fully reveal the complexity and advantage of city-scale networks. Besides, few existing works investigate the mobility patterns evolving of ETs.

\noindent\textbf{Large-scale:}
Different from the existing work, we delve the long-term (over five years) mobility and charging patterns evolving for a large-scale (over 13,000) ET network.
To our best knowledge, ePat is the first work of data-driven investigation on studying the long-term mobility and charging patterns for large-scale ET networks.

\subsection{Summary}

Existing works mostly study the mobility and charging patterns of ETs with small-scale and short-term data, e.g., fewer than 1000 taxis and within one week. Even though some works investigate the mobility patterns of ETs from a large-scale perspective, they only leverage data in several days. However, our ePat is the first work to study the
long-term mobility and charging patterns for large-scale ET networks. Such a long-term data-driven investigation enables us to identify the real-world ET mobility and charging patterns evolving, which cannot be revealed using simulation studies, small-scale data or under a short-term setting.

\begin{figure*}[!hbt] \centering 
\vspace*{-0pt}
\includegraphics[width=0.9\textwidth, keepaspectratio=true]{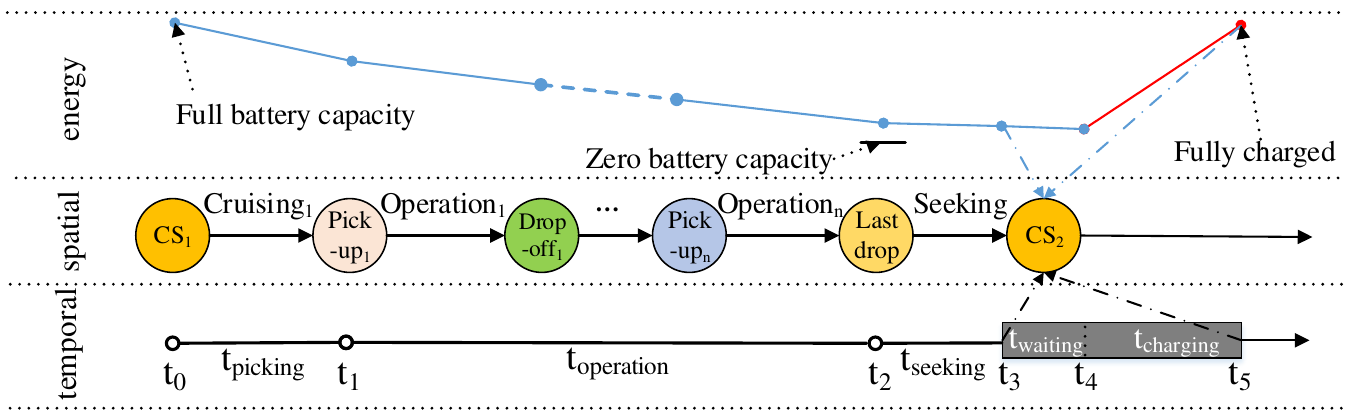}
\caption{Operation and charging patterns of electric taxis}
\label{fig:ETpat}
\end{figure*}

\section{Methodology}
\label{sec:epat}
In this section, we introduce our investigation methodology by
(i) describing our large-scale and multi-source ET datasets;
(ii) contextualizing ET activities by defining operation and charging characteristics; and
(iii) introducing some quantification measurement metrics.

\subsection{Data Description}
\label{sec:data}

For this project, we obtain four datasets from Shenzhen, which is a city with more than 12 million population and an area of 792 mi$^2$.
The time span of these datasets is from September 2013 to July 2018, during which the percentage of ETs among all taxis has increased from 2.7\% to 65.2\%.
The four datasets include
4.8 TB taxi GPS data,
240 GB taxi transaction data,
117 charging station data, and the
urban partition data with 491 regions of Shenzhen.
An example including some primary fields of each dataset is shown in Table~\ref{tab:data}.

\subsection{ET Contextualization}
\label{sec:ocp}

We leverage Figure~\ref{fig:ETpat} to illustrate how we put the operation and charging patterns of ETs into contexts from three dimensions, i.e., spatial, temporal, and energy.
As shown in Figure~\ref{fig:ETpat}, the period from $t_0$ to $t_5$ is defined as a complete operation and charging cycle.

\begin{itemize}

\item \textbf{Overhead After Charging $t_{picking} = t_1-t_0$:}
At time $t_0$, the ET is fully charged and in the full battery capacity status, and then it starts to cruise to seek passengers.
At time $t_1$, it has the first passenger after the charging.
During this process, the battery capacity of the ET is decreasing.
We denote the time from the charging station to pick up the first passenger as $t_{picking}$.

 \item \textbf{Normal Operation $t_{operation} = t_2-t_1$:}
 Due to the limited battery capacity and high range anxiety caused by low battery capacity, ET drivers will start to seek charging stations when the battery capacity declines to a certain level \cite{tian2014understanding}.
 Hence, we define the period from the first pickup after a charge to the last drop-off before the next charge as the $t_{operation}$.
 During this period, the energy level is non-increasing.

\item \textbf{Overhead Before Charging $t_{seeking} = t_3-t_2$:}
After dropping off a passenger at time $t_2$, the ET driver will stop seeking passengers and go to a charging station.
At time $t_3$, it arrives at a charging station, and we denote the time of seeking a charging station as $t_{seeking}$.
Note that an ET driver may not go to the nearest charging station due to various factors \cite{tian2016real}, e.g., traffic conditions, and the $t_{seeking}$ varies when drivers choose different charging stations.
During this period, the energy level is always decreasing.

\item \textbf{Overhead for Charging $t_{waiting} + t_{charging}$:}
At $t_3$, the ET arrives at a charging station.
However, due to the limited charging points and drivers' heuristic charging station searching behaviors, there may be heavy queuing phenomena in the station at $t_3$, which results in long waiting time in this station.
When there is a charging point available in the station at $t_4$, this ET will start to charge and its battery capacity increases.
We leverage the $t_{waiting}$ to stand for the waiting period for an available charging point (i.e., $t_{waiting}=t_4-t_3$), followed by the charging time in the station (i.e., $t_{charging} = t_5-t_4$). After fully charged, this ET will start to cruise and seek for passengers again.
\end{itemize}
\noindent We define
a charging activity as the process of a driver starts to seek a charging station (i.e., $t_2$) and until he/she finishes the charging activity (i.e., $t_5$), which is $t_{seeking}$ + $t_{waiting}$ + $t_{charging}$.
An operation activity includes the overhead after charging $t_{picking}$ because it happens after charging, so the duration of an operation activity is $t_{picking}$ + $t_{operation}$.

\subsection{Quantification Metrics}

\textbf{Mobility Evolving Measurement:}
We investigate the mobility pattern evolving of the ET network from both spatial and temporal perspectives.
For the spatial pattern evolving, we define the ET coverage density and the daily operation distance for quantification.
For the temporal pattern evolving, we utilize the $t_{picking}$ evolving pattern for quantification, which has been defined in Section~\ref{sec:ocp}.

\noindent \textbf{Charging Evolving Measurement:}
For the charging measurement, we investigate the $t_{waiting}$ and $t_{charging}$ evolving patterns. Besides, we define two quantitative metrics, i.e., charging network connectivity and charging station utilization rate to quantify the charging network evolving. We also investigate the charging supply and demand evolving patterns from both temporal and spatial aspects over five years, from which some insights are derived.

\begin{figure*}[htb] \centering
\vspace*{-0pt}
\includegraphics[width=0.98\textwidth, keepaspectratio=true]{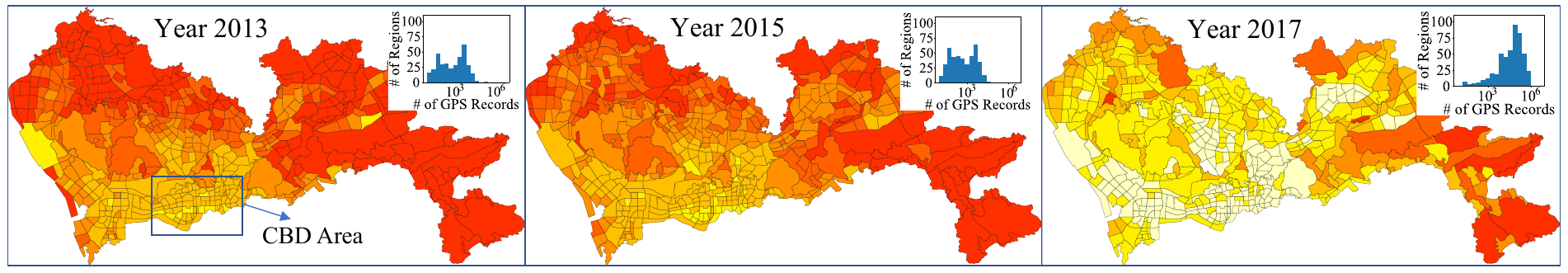}
\vspace*{-0pt}
\caption{The ET coverage density evolving pattern in 491 urban regions}
\vspace*{-0pt}
\label{fig:visual1}
\end{figure*}

\section{Investigation Results}
\label{sec:result}
We present our measurement results from four aspects:
(i) the ET mobility evolving patterns;
(ii) the ET charging evolving patterns;
(iii) the impacts of different contexts; and
(iv) the benefits of ET evolving. The details are shown below.

\subsection{Mobility Evolving}
\noindent \textbf{ET Coverage Density Evolving:}
We define the ET coverage density as the number of GPS in a specific region to investigate the ET mobility patterns.
We leverage the urban partition data to divide Shenzhen into 491 regions,
which is provided by the Shenzhen government.
As in Figure~\ref{fig:visual1}, the darker red means fewer ET activities in this region and the lighter yellow stands for more ET activities in this region.
We also show the number of GPS records distribution in these regions in the northeast corners.
We have the following observations:
(i) The ET density in Shenzhen has increased significantly during the five years, especially from 2015 to 2017, almost all the regions have a higher ET density compared to the previous year, which was also observed from the distributions.
Quantitatively, the growth of the density from 2015 to 2017 has increased 25 times compared with the previous growth;
(ii) the ETs are gathered in the central business district (CBD) area from 2013 to 2015, but more ETs start to operate in the suburban areas after 2015.
One factor may be the urbanization process in Shenzhen, e.g., more companies built and more people live in former suburban areas \cite{chang2013spatial}.
Another reason may be the upgrading of the charging infrastructures, e.g., more charging stations are built in suburban areas, which we will investigate in Section~\ref{sec:station}.

\begin{figure}[!htb]
\vspace*{0pt}
\begin{minipage}[c]{0.235\textwidth} \centering
\vspace*{-2pt}
\hspace*{-0pt}
\includegraphics[width=1.0\textwidth, keepaspectratio=true]{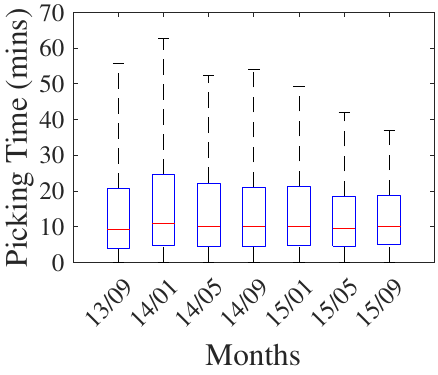}
\vspace*{-16pt}
\caption{$T_{picking}$ evolving} \label{fig:tpick}
\vspace*{-5pt}
\end{minipage}
\begin{minipage}[c]{0.235\textwidth} \centering
\vspace*{-2pt}
\hspace*{-2pt}
\includegraphics[width=1.0\textwidth, keepaspectratio=true]{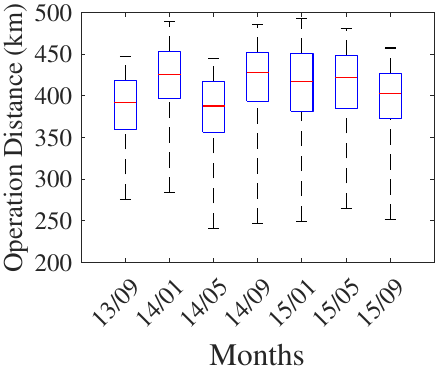}
\vspace*{-16pt}
\caption{Op. dist. evolving} \label{fig:OpDist}
\vspace*{-5pt}
\end{minipage}
\vspace*{3pt}
\end{figure}

\textbf{Operation Activity Evolving:}
In this subsection, we investigate the long-term operation evolving patterns of ETs using $t_{picking}$ and daily operation distance, which are shown by box plots as Figure~\ref{fig:tpick} and Figure~\ref{fig:OpDist}.
The top and bottom of each ``box'' are the 25th and 75th percentiles of the data.
The middle red lines are the median values.
The top and bottom of each black dash line indicate the maximal and minimal values.
The top red crosses are outliers, which are values that more than 1.5 times the interquartile range away from the top value or bottom value of the box.

 Figure~\ref{fig:tpick} shows the evolving of $t_{picking}$,
i.e., the duration from the fully charged time to the first pickup time as shown in Figure~\ref{fig:ETpat}.
We found that the median value of $t_{picking}$ is almost stable at 10 minutes,
while the maximal values have a trend of decrease.
We show the daily operation distance of ETs in Figure~\ref{fig:OpDist}.
Intuitively, there are no obvious evolving patterns since the new adopted ETs can increase the average operation distance of the ET fleet even though the battery degradation nature, which results in relatively stable average daily operation distance.
However, there is a significant drop in May 2014. We carefully studied the causality, and we found there is exceptionally severe weather in this month with heavy rain, resulting in many ETs break down.

\subsection{Charging Evolving}
\label{sec:station}
We perform an in-depth investigation of the charging activity evolving for ETs, coupled with the evolving patterns of the accessory charging network.

\subsubsection{Charging Activity Evolving}
We leverage $t_{waiting}$ and $t_{charging}$ to investigate the charging activity evolving of ETs from year 2013 to 2017.

\begin{figure}[!htb]
\begin{minipage}[c]{0.235\textwidth} \centering
\hspace*{-0pt}
\includegraphics[width=1.0\textwidth, keepaspectratio=true]{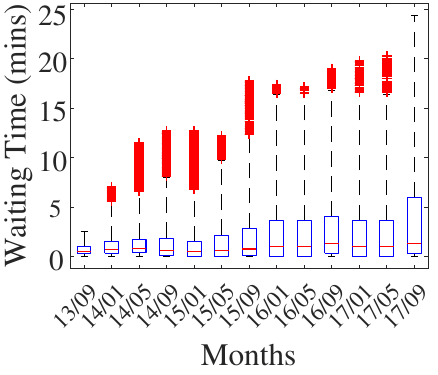}
\vspace*{-15pt}
\caption{$T_{waiting}$ evolving} \label{fig:Waiting}
\vspace*{-6pt}
\end{minipage}
\begin{minipage}[c]{0.235\textwidth} \centering
\includegraphics[width=1.0\textwidth, keepaspectratio=true]{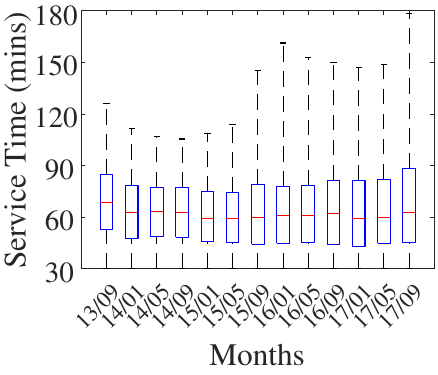}
\vspace*{-15pt}
\caption{$T_{charging}$ evolving} \label{fig:service}
\vspace*{-6pt}
\end{minipage}
\end{figure}

As shown in Figure~\ref{fig:Waiting}, the overall trend for $t_{waiting}$ is growing from 2013 to 2017.
We then further investigate causalities behind this phenomenon and one possible reason is that
the number of ETs increased too fast;
whereas the increase of charging infrastructure cannot keep up.
Specifically,
from 2013 to 2014, the number of ETs had increased about two hundred,
but the number of charging points only increased by 2, which causes waiting time increases.
From 2015 to 2017, the number of ETs had dramatically increased;
whereas only limited charging stations are built.

Figure~\ref{fig:service} shows the average $t_{charging}$ evolving pattern. Even though the highest charging time has an increasing trend, the medians, the 25th, and the 75th percentiles are close, i.e., about 60, 45, and 75 minutes, respectively.
It indicates that most drivers have a relatively stable charging time.
Since the $t_{seeking}$ is mainly related to the urban traffic conditions,
which is decided by many non-mobility factors, e.g., government traffic management and urban road network evolving, we do not report it as it is difficult to find obvious evolving patterns even after an in-depth investigation.

\begin{figure}[!htb] \centering
\vspace*{3pt}
\includegraphics[width=0.48\textwidth, keepaspectratio=true]{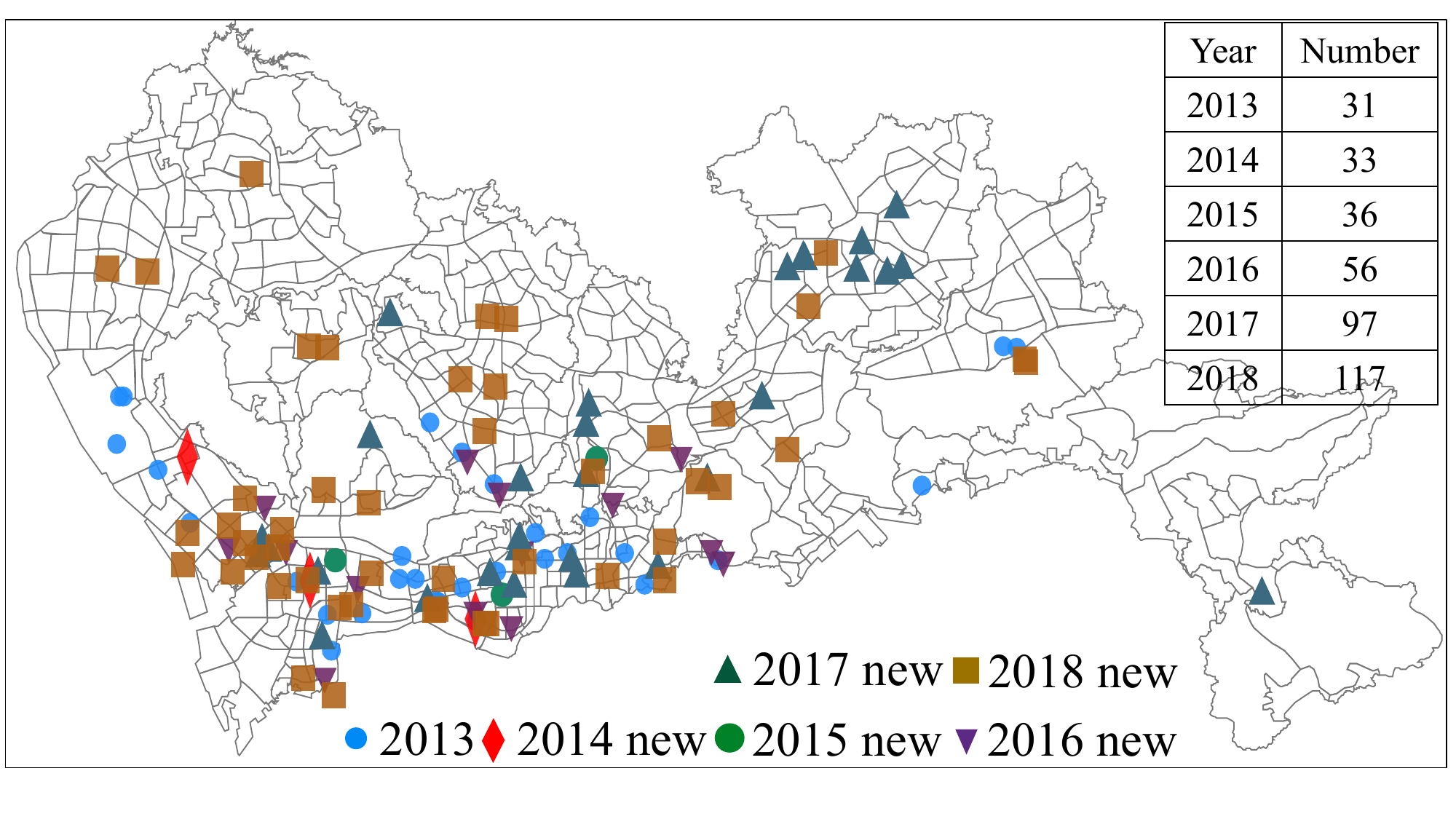}
\vspace*{-15pt}
\caption{Charging station deployment evolving}
\vspace*{-5pt}
\label{fig:stationvis}
\end{figure}


\subsubsection{Charging Supply Evolving}
\label{sec:station}
\begin{figure}[!htb]
\begin{minipage}[c]{0.235\textwidth} \centering
\vspace*{2pt}
\hspace*{-0pt}
\includegraphics[width=1.0\textwidth, keepaspectratio=true]{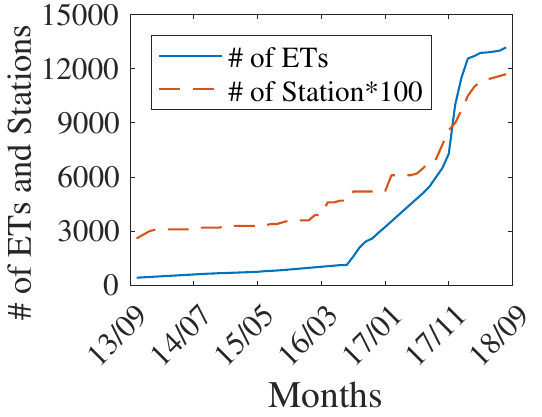}
\vspace*{-13pt}
\caption{ETs \& charging stations increase} \label{fig:ETNum}
\vspace*{-6pt}
\end{minipage}
\begin{minipage}[c]{0.235\textwidth} \centering
\vspace*{2pt}
\hspace*{-0pt}
\includegraphics[width=1.0\textwidth, keepaspectratio=true]{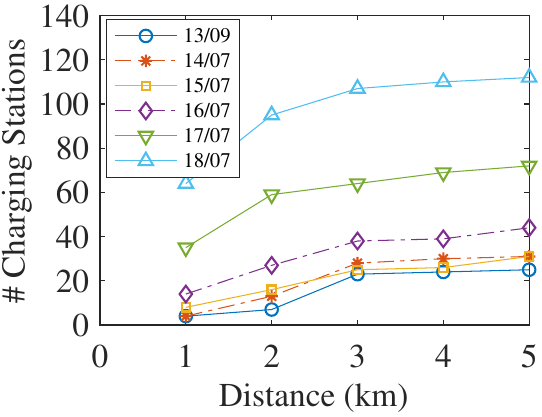}
\vspace*{-13pt}
\caption{Charging station distance evolving} \label{fig:csdistribution1}
\vspace*{-6pt}
\end{minipage}
\end{figure}

In Figure~\ref{fig:stationvis},
we visualize the evolving pattern of charging supply for ETs in Shenzhen from year 2013 to 2018 in 491 urban regions.
We found the total number of charging stations has increased from 31 in 2013 to 117 in 2018. Intuitively, ET charging stations are unevenly distributed in Shenzhen,
and the general trend of the charging station deployment is from the urban areas to suburban areas, with an accelerated deployment speed from 2013 to 2018, especially from 2016 to 2018, during which more than 60 charging stations were deployed.
We further compare the increase of charging stations and ETs, which is given in Figure~\ref{fig:ETNum}.
We found that the number of ETs has not increased too much from September 2013 to June 2016, but there is a large-scale adoption after 2016.
Especially at the end of 2017, there is a sharp expansion of ETs, while the growth rate of charging stations is much lower.

We further investigate the quantitative evolving patterns of the charging supply by defining the charging network connectivity, which is the average shortest distance between charging stations and their neighbor stations shown as Equation~\ref{eq:connectivity}.
\vspace*{-13pt}
\begin{eqnarray}
Conn(net) = \frac{{\sum\limits_{i = 1}^N {S{D_i}} }}{N}
\vspace*{-16pt}
\label{eq:connectivity}
\end{eqnarray}
\noindent where
$N$ is the total number of charging stations in the charging network $net$;
$SD_i$ is the shortest distance neighbor of the $i$th charging station.

The charging network connectivity can reflect on average how far there is another charging station for a given charging station.
We first study the distribution of the distances between any two charging stations.
We found that the distances between two charging stations can be as short as 200 meters and as long as 70 km based on our statistics, but most charging stations have a neighbor station within 5 km.
Figure~\ref{fig:csdistribution1} shows the number of charging stations with at least another charging station within a certain distance, e.g., from 1 km to 5 km over five years.
From Table~\ref{tab:conn}, we found the charging network connectivity has increased from 2.76 in 2014 to 1.31 in 2018, which indicates the charging network in 2018 is more resilient to cope with the long waiting phenomena in charging stations.
We also found that in recent two years most charging stations (e.g., 96 stations in 2018, i.e., more than 82\% of all 117 stations) have at least one station within 2 km, making the Shenzhen ET charging network well connected.
In this case, if some drivers find there are no charging points available when they arrive at a station, they can quickly go to other nearby stations.

\begin{table}[h!]\small
\vspace*{-5pt}
\caption{Charging network connectivity evolving}
\vspace*{-5pt}
\centering
\begin{tabular}{ |c|c|c|c|c|c|c|}
\hline
Year & 2013 & 2014&2015&2016&2017&2018\\
\hline
Connectivity (km)& 2.25 & 2.76 &2.62& 2.13&2.00&1.31 \\
\hline
\end{tabular}
\vspace*{-12pt}
\label{tab:conn}
\end{table}

\subsubsection{Charging Demand Evolving}
\begin{figure}[!htb]
\begin{minipage}[c]{0.235\textwidth} \centering
\vspace*{0pt}
\hspace*{-0pt}
\includegraphics[width=1.0\textwidth, keepaspectratio=true]{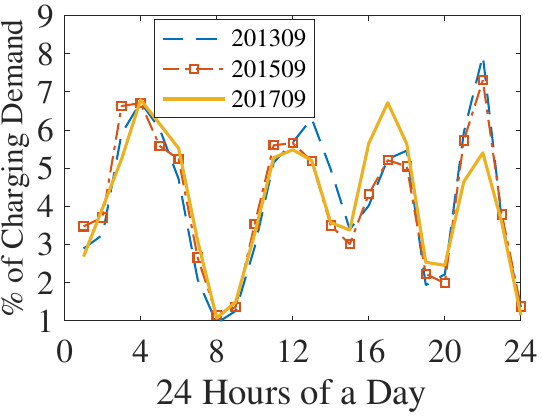}
\vspace*{-11pt}
\caption{Charging temporal pattern} \label{fig:temporal1}
\vspace*{-5pt}
\end{minipage}
\begin{minipage}[c]{0.235\textwidth} \centering
\vspace*{0pt}
\hspace*{-0pt}
\includegraphics[width=1.0\textwidth, keepaspectratio=true]{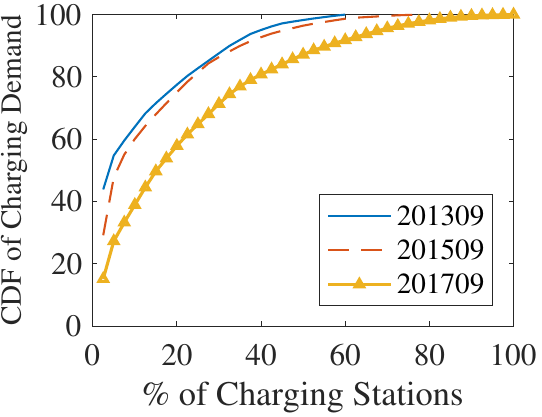}
\vspace*{-11pt}
\caption{Charging spatial pattern} \label{fig:spatial_cdf1}
\vspace*{-5pt}
\end{minipage}
\end{figure}

To understand the charging demand evolving of ETs,
we investigate both the spatial and temporal charging demands of the ET network, which are shown in Figure~\ref{fig:temporal1} and Figure~\ref{fig:spatial_cdf1}, respectively.
As shown in Figure~\ref{fig:temporal1}, we found there are four charging demand peaks in each day, i.e., 3:00-5:00, 11:00-13:00, 16:00-17:00 and 22:00-23:00.
Such a temporal pattern has not changed significantly in the last five years.
Among the four charging peaks, the first and third ones are before the rush hours for picking up taxi passengers, which can make them have enough battery energy to pick up passengers. The second one is during the lunchtime when the electricity price is relatively low. We then investigate the electricity prices during the four peaks, and we find all of the four peaks are in relatively low electricity price durations.

Figure~\ref{fig:spatial_cdf1} shows the accumulative charging demand in the charging network. Further, we found out that 20\% of charging stations accommodate 80\% of charging demand in 2013 due to their convenient locations, e.g., most of them are located in downtown areas, where the possibility for ETs to pick up passengers after charging is high.
However, such an unbalanced spatial pattern has improved since 2015 given the increased charging stations. For example, the charging demand in top 20\% popular charging stations accounts for only 60\% of the total charging demand of the ET fleet in September 2017, which indicates the unbalanced charging station utilization phenomenon has improved with more charging stations deployed.

\subsubsection{Charging Station Utilization Rate Evolving}
To further study the relationship between charging demand and supply at the station level,
we define the utilization rate of a charging station $i$ as the ratio of the daily number of charging events $CE(i)$ over the charging points $CP(i)$ in this station, as shown in Equation~\ref{eq:ur}.

\vspace*{-14pt}
\begin{eqnarray}
UR(i) = \frac{{CE(i)}}{{CP(i)}}
\vspace*{-14pt}
\label{eq:ur}
 \end{eqnarray}

\begin{figure}[htb] \centering
\vspace*{-2pt}
\includegraphics[width=0.47\textwidth, keepaspectratio=true]{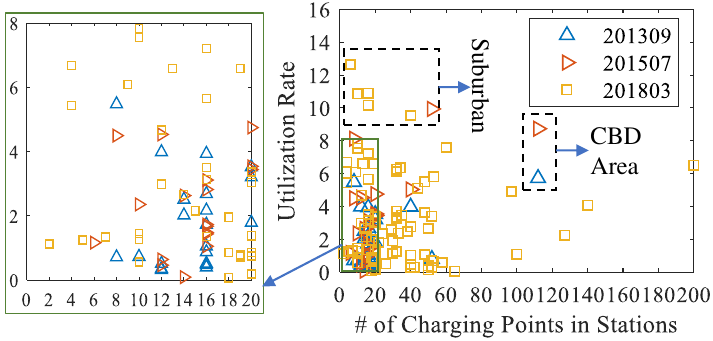}
\vspace*{-3pt}
\caption{Utilization rate evolving}
\vspace*{-2pt}
\label{fig:spatial_ura}
\end{figure}

Figure~\ref{fig:spatial_ura} shows the charging station utilization evolving during the five years.
We found that the highest utilization rate has doubled from 2013 to 2018 as more ETs promoted, resulting in more charging demand.
However, the charging station sizes with the highest utilization rates are different, so we further investigate the geographical feature of the charging stations.
We found the highest utilization rate of 2013 happens in the largest charging station in the CBD area, which has 112 charging points.
This station was also the largest charging station in 2015 and it remained relatively high utilization rate in 2015.
However, it has been closed in 2016 due to some security concerns.
In 2018, there are five large charging stations with more than 100 charging points located in Shenzhen suburban areas to accommodate the skyrocketing ETs.
Even though the utilization rates in these charging stations is not too low, they are not too high compared to some small charging stations.
We found that the highest top 5 utilization rates' charging stations have the same common features, i.e., (i) located in suburban areas, e.g., Shenzhen airport, and (ii) include less than 40 charging points, which is an interesting finding.

The above utilization rate evolving patterns have the potential to provide guideline for other cities, which have plans to promote large-scale ETs.
For example, if it is feasible to deploy large charging stations in the urban CBD areas, then those charging stations have the potential to achieve the highest utilization rate, which can improve the charging infrastructure utilization and reduce the resource waste.
If it is not feasible, more median/small charging stations with less than 40 charging points should be placed in some important suburban areas, e.g., airports and transportation hubs.

\subsubsection{Charging Station Field Study}
\label{sec:field}
\begin{figure}[htb] \centering
\vspace*{-2pt}
\includegraphics[width=0.47\textwidth, keepaspectratio=true]{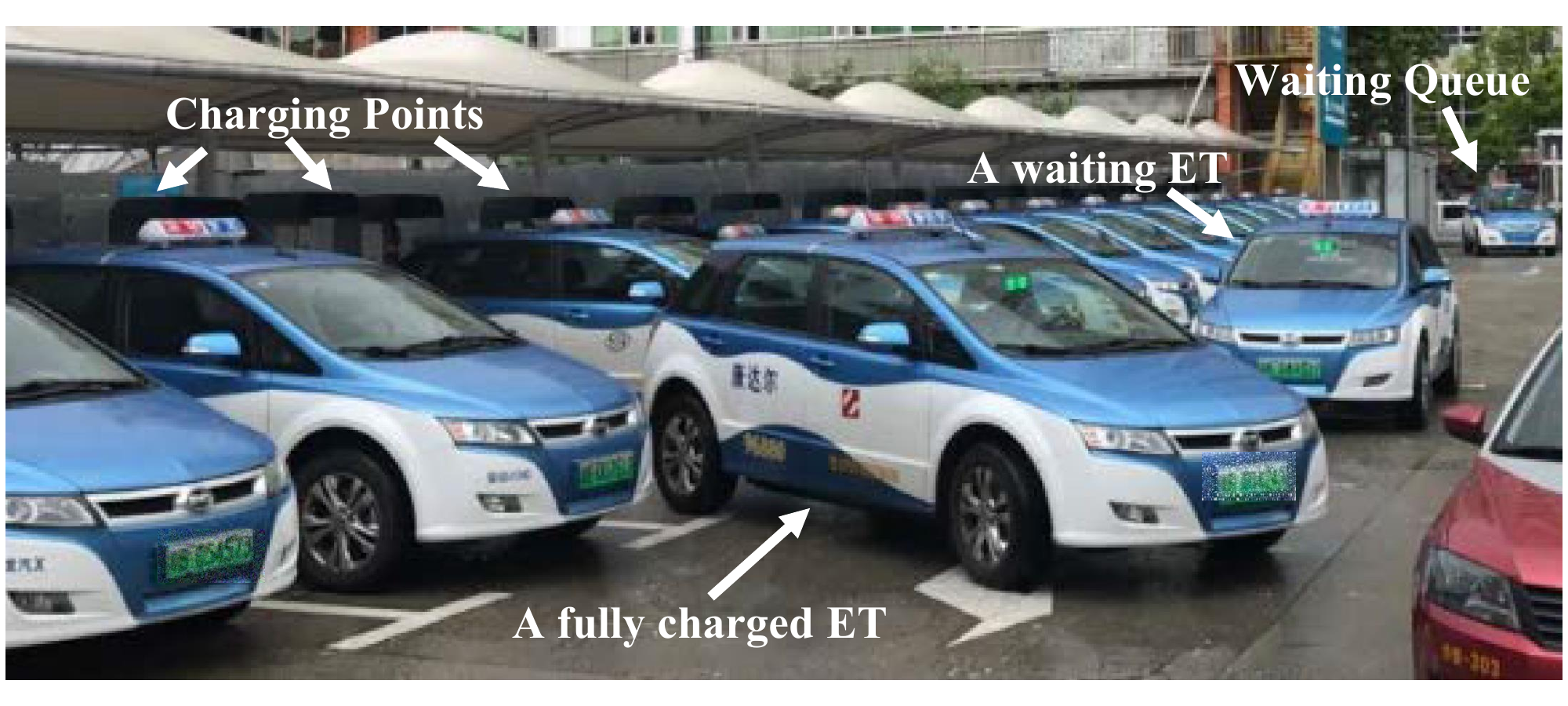}
\vspace*{-2pt}
\caption{Charging and queuing in our field study}
\vspace*{-2pt}
\label{fig:fieldstudy1}
\end{figure}

To investigate the charging patterns of ETs in Shenzhen, our team has performed a series of field studies at charging stations in Shenzhen from 2015 to 2017. Figure~\ref{fig:fieldstudy1} shows the charging station status observed by us during one of the field studies.
As we can see,
a line of charging points are installed under a shelter, which is used for charging protection;
an ET was queuing at the entrance of the station since there were no charging points available at that time (12:31 PM);
another ET was approaching a charging point since there was a fully charged ET leaving the charging point.
In this field study, we found all of the 42 charging points in this station were occupied from 11:30 - 13:00 and there were queuing phenomena since 12:26 PM and the longest waiting time is up to 31 minutes, which indicate this charging station may have a high utilization rate and a severe queuing phenomenon. This long-time queuing phenomenon also reflects the unbalanced charging demand and/or insufficient charging supply.

\subsection{Contextual Factors}

\subsubsection{Impact of Daytime and Nighttime}

\begin{figure}[!htb]
\begin{minipage}[c]{0.235\textwidth} \centering
\vspace*{2pt}
\hspace*{-0pt}
\includegraphics[width=1.0\textwidth, keepaspectratio=true]{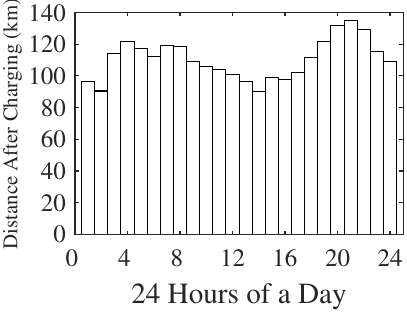}
\vspace*{-10pt}
\caption{After-charging distance} \label{fig:daynight}
\vspace*{-6pt}
\end{minipage}
\begin{minipage}[c]{0.235\textwidth} \centering
\vspace*{-0pt}
\hspace*{-0pt}
\includegraphics[width=1.0\textwidth, keepaspectratio=true]{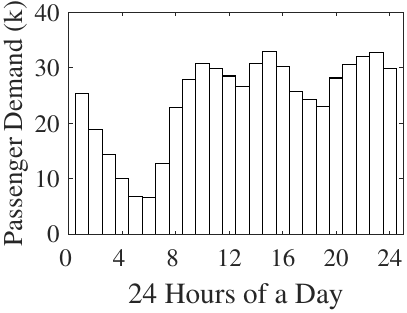}
\vspace*{-10pt}
\caption{Passengers' travel demand} \label{fig:Pssen}
\vspace*{-6pt}
\end{minipage}
\end{figure}

In Figure~\ref{fig:daynight}, we found that the operation distance of ETs is different if they charged in different hours.
For example, if an ET is charged at 3:00, it may be fully charged at 4:30 and leave, then it travels about 120 km before next charging.
We also found that the charging starting time from 3:00-8:00 and 20:00-22:00 may result in a more extended operation distance for ETs even with potential congestion at rush hours. We then explore the possible reason by considering the passengers' travel demand, as shown in
Figure~\ref{fig:Pssen}. We found that there is higher taxi demand around 9:00 and 22:00,
which is alighted with the period that ETs were fully charged and started to pick up passengers.
It may indicate that ET drivers would prefer to charge before the rush hour for longer cruising distances, which can potentially reduce their income loss caused by the long charging time.

\subsubsection{Impact of Temperatures \& Weather}
\label{sec:temp}
\vspace*{2pt}
We select three temperature levels and three weather conditions to investigate their impacts on ETs' mobility patterns, i.e.,
high (27 - 34 $^{\circ}$C),
low (6 - 8 $^{\circ}$C),
mild (17 - 25 $^{\circ}$C), and sunny, heavy rain and typhoon.
As in Table~\ref{tab:context},
the average operation distance under mild temperature is the longest;
the average operation distance under low-temperature is the shortest.
One explanation is that the energy consumption rate is the highest in cold days because of the air conditioners for heating.

As the impacts of different weather,
we found that the average operation distance of ETs in sunny days is longer than rainy days; the operation distance in hurricane days is the shortest.

The potential reason could be that in sunny days, ET drivers may reduce the charging time for longer operation distance.
In rainy days, drivers have longer operation distance than in typhoon days.
Hence, we conclude that different weather conditions have various impacts on the mobility patterns of ETs, e.g., severe weather conditions will potentially reduce the daily operation distances of ETs.

\begin{table}[h!]
\small
\vspace*{-2pt}
\caption{Daily Oper. Distance in Different Contexts}
\vspace*{-1pt}
\centering
\begin{tabular}{ |c|c|c|c|}
\hline
Weather & Distance (km) & Temperatures & Distance (km)\\
\hline
Sunny & 447 & Mild & 437\\
\hline
Heavy Rain & 413 & High & 421\\
\hline
Typhoon & 406 & Low & 389\\
\hline
\end{tabular}
\vspace*{-0pt}
\label{tab:context}
\end{table}

\vspace*{-0pt}
\subsection{ET Benefits}
Using the emissions and driver incomes as examples,
we quantificationally compare the ETs with conventional gas taxis to study the evolving of ETs' benefits based on their daily mobility patterns, e.g., operation distances.
\subsubsection{Emission Reduction}

Based on \cite{oguchi2002carbondioxide}, we consider the real-world traffic conditions (e.g., travel speed), the daily travel distance of ETs, and daily travel time of ETs to accurately estimate CO$_2$ emission reduction of the ET network, which is represented by
\begin{eqnarray}
E=K_c*\left [ 0.3T+0.028D+0.056\sum_{i=1}^{k}(\delta_i*(v_{i+1}^{2}-v_{i}^{2})) \right ]
\label{eq:emission}
\end{eqnarray}

\noindent where $E$ is the CO$_2$ emissions (g);
$T$ is the travel time of taxis (s);
$D$ is the travel distance of taxis (m);
$K_c$ is the coefficient between gasoline consume and CO$_2$ emissions, which is 2322g (CO$_2$)/liter(gasoline) for Shenzhen taxis;
$v_i$ is the speed at time $i$ (m/s);
$\delta_i$ is a tow-value indicator, taking the value 1 when accelerating ($v_{i+1} > v_i$) otherwise 0 ($v_{i+1} \leq v_i$).
\begin{wrapfigure}{r}{0.26\textwidth}
\includegraphics[width=0.23\textwidth, keepaspectratio=true]{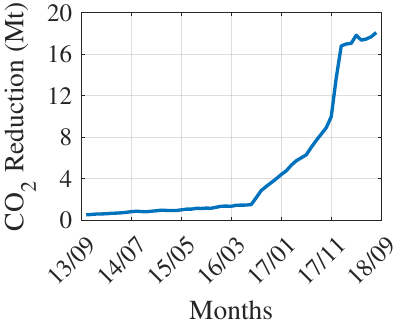}
\vspace*{-5pt}
\caption{CO$_2$ reduction evolving} \label{fig:CO}
\vspace*{-8pt}
\end{wrapfigure}
Thus, we estimate the monthly CO$_2$ reduction from September 2013 to July 2018 in Figure~\ref{fig:CO}.
There is more ETs deployment from June 2016, which explain a major CO$_2$ reduction.
Especially in the last two months of 2017, the number of ET has increased significantly, e.g., in December 2017, the CO$_2$ reduction is close to 1.8 million tons, which is equivalent to CO$_2$ emissions from 176,324 homes' energy use for one year according to the U.S. Environmental Protection Agency.

\subsubsection{Driver Income Comparison}
A key roadblock of ET deployment is that many drivers worry about their income reduction caused by the long charging time of ETs compared with refueling of conventional gas taxis.
Here, we comprehensively consider the daily travel distance of taxis to calculate the profits of ETs and gas taxis, which can be represented by
\vspace*{2pt}
\begin{eqnarray}
P(j) = I(j) - E(j)
= I(j) - D(j)*C_r*P
\label{eq:profit}
\end{eqnarray}
\vspace*{2pt}
\noindent where $P(j)$ and $I(j)$ stand for the daily profit and income of driver $j$. $E(j)$ includes all costs, e.g., refueling costs, maintenance fees.
In this paper, we only consider the refueling costs because they are the major expenses for the daily operation of taxis, which depend on taxis' daily operation distance $D(j)$ and energy consumption rate $C_r$.
For a gas taxi, the fuel consumption per 100 km is 9 liters,  so the energy consumption rate $C_r$ for gas taxis is 9 liter/100km. Besides, the unit oil price $P$ is 1.015\$/liter (2017) in Shenzhen.
For ETs, the energy consumption rate $C_r$ is 26kWh/100km and the average electricity price $P$ is 0.1\$/kWh in Shenzhen.

\begin{figure}[!htb]
\begin{minipage}[c]{0.235\textwidth} \centering
\vspace*{-5pt}
\includegraphics[width=1.0\textwidth, keepaspectratio=true]{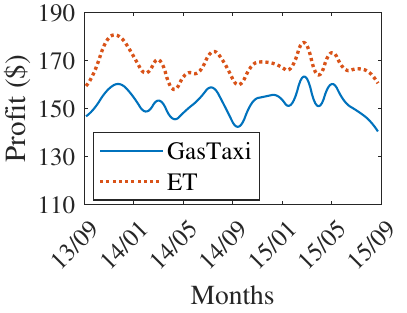}
\vspace*{-12pt}
\caption{Daily Profit comparison} \label{fig:Income}
\vspace*{-6pt}
\end{minipage}
\begin{minipage}[c]{0.235\textwidth} \centering
\vspace*{-5pt}
\includegraphics[width=1.0\textwidth, keepaspectratio=true]{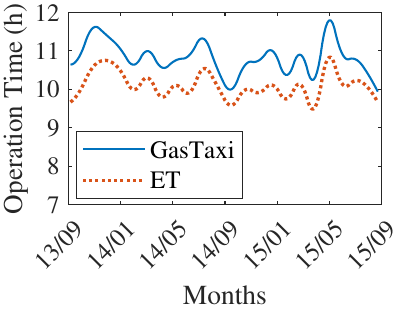}
\vspace*{-12pt}
\caption{Daily operation time comparison} \label{fig:TimeReduction}
\vspace*{-6pt}
\end{minipage}
\end{figure}

In Figure~\ref{fig:Income}, we show the profits evolving of ET drivers and gas taxi drivers during two years.
We found that the average daily income of each ET driver is about 15\$ more than each gas taxi driver per day even though the ET drivers have lower income.
Besides, we also found that the daily operation time of ETs is about 5\% less than gas taxis, from 1.1 h to 0.26 h, which is shown in Figure~\ref{fig:TimeReduction}.
When considering these two figures, we found that even though the operation time decreases for ETs,
the profit of ETs can potentially increase by energy saving, in addition to other benefits, i.e., fewer emissions, less noise, and more rest time for drivers during charging.
However, since our work focused on operation of ETs, our analyses did not account for the initial vehicle purchase cost of ETs, which are typically higher than gas taxis. For example, even with Shenzhen government subsidies, a BYD e6 costs \$29,400; whereas a gas taxi costs \$13,200.

\section{Insights and Discussions}
\label{sec:insight}
In this section, we provide a few insights we obtained in our measurement study and some discussions.

\subsection{Insights}
Our long-term investigation has the potential to provide some insights for (i) city governments, e.g., how to deploy charging stations and address the unbalanced charging supply and demand issues; (ii) taxi drivers, e.g., their profits would not reduce although they have more time to rest; and (iii) society, e.g., how to promote large-scale ET networks and how much the quantitative emission reduction. We summarize some key insights below.

\textbf{Charging Station Deployment:}
Charging stations play a key role in large-scale ET promotion. Insufficient charging infrastructure will result in a longer waiting time for charging points, as shown in Figure~\ref{fig:Waiting}. However, even though enough charging points are deployed, the unbalanced temporal or spatial charging demand and supply reduce the efficiency of the overall charging network.
We found that the uneven supply and demand phenomena can be improved by building more charging station in the right locations, as shown in Figure~\ref{fig:spatial_cdf1}.
Based on our investigation of Figure~\ref{fig:spatial_ura}, we found that those large charging stations in urban CBD area has the best chance to achieve a higher utilization rate, which can improve the charging infrastructure utilization and reduce the charging resource waste. \textit{Hence, if it is feasible to deploy large stations in urban CBD area for other cities, it would be beneficial for promoting their large-scale ET networks; if it is hard to place large stations in CBD areas due to some reasons (e.g., land resources unavailable), more median/small charging stations with less than 40 charging points may be placed in some important suburban areas (e.g., airports and transportation hubs), to improve the overall charging efficiency.}

\textbf{Charging Behavior Patterns:} There is a stable temporal charging pattern of ET drivers as shown in Figure~\ref{fig:temporal1}. We found a very large gap between the four charging peaks and the charging valleys, which inevitably result in the underutilized charging and the overcrowded charging during different hours. The major causality behind this phenomenon is that Shenzhen has \textit{time-varying electricity prices}, and the charging peaks happen in relatively low price periods. Another reason is that the ET drivers may have the potential to pick up more passengers as some charging peaks are one hour before the rush hours. Once this phenomenon exists, it must need to build more charging stations to accommodate ETs, while it will also cause the resource waste in many hours of every day. This phenomenon will happen with a very high possibility during the ET promotion process of other cities, and it is important to address it. \textit{One possible approach is to adjust the electricity prices, e.g., setting lower electricity prices during the valleys and higher prices for the peaks for ETs only. For taxi drivers, our investigation can guide them to charge for minimizing their $t_{waiting}$ and reduce the possible operation loss, e.g., they can charge during 23:00-3:00 as there is the lowest electricity price in Shenzhen.}

\textbf{Charging Contexts:} The following three contexts are explored for their impacts on ET mobility patterns.
\begin{itemize}
  \item \textbf{Nighttime:}
 \textit{There is no obvious observation indicates that the cruising mileage of ETs decreases significantly during the night} as shown in Figure~\ref{fig:daynight},
 even though we thought that the cruising mileage should drop because of the headlights compared to the daytime.
 The reason may be that typically there is no server traffic congestion during nights, which results in more extended cruising range compared to the daytime even with energy consumption of headlights \cite{SZET5}.
 \item \textbf{Temperature and Weather:}
Even though the adverse temperatures and weathers will decrease the daily operation distance of ETs, e.g., about 4\% in hot days and 9\% in hurricane days, the reduction is still less than operation distance reduction of gas taxis under the same situations based on our investigation. We do not report the details of gas taxis because we focus on the ET patterns in this paper. \textit{This finding can help better promote ETs, since the ETs may achieve a higher energy efficiency even in adverse contexts.}
\end{itemize}

\textbf{ET Benefits:} Even though the overall daily operation time of ETs has reduced compared to gas taxis, the profits of ET drivers are potentially higher than gas taxis' due to the low electricity prices, (e.g., ET drivers' daily profits are about \$20 higher than gas taxi drivers'). This finding may lead to a higher incentive for drivers to accept electric vehicles. Most importantly, the overall emission reduction can be beneficial for public health and air quality (e.g., CO$_2$ reduction by adopting ETs in Shenzhen is about 1.8 million tons in December 2017), which is important for a sustainable society.

\subsection{Discussions}
\noindent\textbf{Limitation.}
In this work, we only utilize ET data from Shenzhen to study the ET evolving process. Due to certain features of Shenzhen (e.g., the most crowded city in China with the fastest economic growth, a tropical city with mild winters), the results we have in Shenzhen may not be applied all other cities. However, we argue that our investigation along with electric vehicle operation models \cite{li2015growing,DongREC,liu2016optimal,kong2016line} can provide envisions for other cities to predict and understand their ET evolving process by considering their own city features.

\textbf{Privacy Protections and Data Management.} During this project,
we establish a secure and reliable transmission mechanism with a wired connection, which feeds our server the filtered ET data collected by Shenzhen transportation committee by using a cellular network.
The filtering process replaces sensitive data, e.g., plate ID with a serial number for privacy protections.
We utilize a 34 TB Hadoop Distributed File System (HDFS) on a cluster consisting of 11 nodes, each of which is equipped with 32 cores and 32 GB RAM.
For daily management and processing, we utilize the MapReduce based Pig and Hive since our analyses are based on log data, instead of streaming data.
Due to long-term GPS data and transaction data, we have been dealing with several kinds of errant data (e.g., duplicated data, missing data, and logical errors of data) to obtain the data for this project.

\textbf{A Balance between ET and Station Evolving:} The relationship between electric vehicles and charging infrastructure (i.e., both charging stations and charging points) is interdependent on lots of factors. If the charging points and charging stations are not enough, it is challenging to promote electric vehicles. However, if the charging infrastructure is much more than necessary in a city with only a small number of electric vehicles, many charging points will be idle, leading to low utilization rates. As a result, a balance needs to be achieved between the number of electric vehicles and the number of charging stations and charging points. \textit{One possible solution to achieve the balance could be designing an intelligent charging recommendation system for the ET network under a centralized management mode. In this case, we can know information of all ETs and charging stations in the ET network, and then we can recommend ETs to charging stations for minimizing their charging overhead and balancing the utilization rates of charging stations at the same time. With the charging recommendation system, we can potentially know how many charging stations/points are enough for the ETs to avoid low-efficient charging resource deployment.}

\textbf{Current Impacts.}
Based on ePat, our understanding of mobility and charging patterns of ET fleets at city scale can be valuable to charging infrastructure providers, ET drivers and government managers. For example, a charging infrastructure provider can build more efficiently charging stations near the frequently-traveled road segments by ETs. Further, taxi drivers may also potentially reduce their charging overhead because there are more charging points available in short distance. Moreover, Shenzhen government agency will benefit from mobility and charging pattern understanding as well because they may evaluate their current ET development for better future upgrading. We have reported our insights from our investigation to Shenzhen transportation committee for better charging station deployment and ET adoption, including some recommended charging station locations and the possible electricity price mechanism for ET charging. These insights were well received but it needs to take some time to see the actual results. Even though the government officials think our results provide valuable technical insights, the real-world deployment is extremely complicated and mostly dependent on policies.
As a result, instead of deciding the development and evolving strategy, our next objective is to utilize our framework to quantify the benefit of a pre-defined evolving strategy.

\noindent \textbf{Potential Impacts.}
(i) A good charging station deployment strategy is also beneficial for promoting large-scale ET networks. Our investigation has the potential to guide other cities to deploy the charging stations more efficiently and economically, achieving success ET promotion. For example, We can guide them where to deploy charging stations and how many charging points are appropriate for each charging station. More specifically, urban CBD areas would be the best places for large cities if the land resources are available for them. Some median/small charging stations with less than 40 charging points in some important suburban areas are also good choices for charging station deployment. We can also estimate how many charging points are sufficient for other cities based on Shenzhen model and their taxi operation patterns. (ii) Our investigation can help people learn the quantitative benefits of ETs and accept this new thing, which can potentially promote other electric vehicles, e.g., electric cars. (iii) Based on the current prediction \cite{kang2017autonomous}, shared autonomous vehicles \cite{chen2016operations, krueger2016preferences, fagnant2018dynamic, johns2016exploring, chen2016management, dandl2018comparing} are most likely to be electric vehicles, which will be cursing around a city to pick up and drop off passengers or parking until dispatched. These features make shared autonomous vehicles very similar to ETs, except the human driver factors. Hence, our investigation on ET evolving process is extremely valuable to predict and quantify the impact of future shared autonomous vehicles. For example, our results on various metrics may be reapplied to shared autonomous vehicles with some modified targeting human driver factors, future parking, and charging models, etc.

\vspace*{-5pt}
\section{Conclusion}
\label{sec:conclusion}
In this paper, we conduct the first work to understand the long-term ET network evolving mobility and charging patterns based on the ET network in the Chinese city Shenzhen, which includes more than 13,000 ETs and 117 charging stations.
Our study is from a comprehensive spatiotemporal perspective to investigate the long-term evolving patterns of an ET network.
We provide a few insights regarding the evolving process of charging
station deployment and utilization, ET coverage density, operation, and mobility patterns under various contexts.
We also quantify the emission reduction and driver income benefits of the ET network based on their mobility patterns.
For the immediate benefit, understanding long-term evolving patterns of Shenzhen ET network provides valuable experiences for Shenzhen and other cities to further promote ETs. For the long-term benefit, our ePat may be used to predict the evolving process of future shared autonomous vehicles and quantify their benefits due to their similar characteristics to ETs.

\vspace*{-5pt}
\begin{acks}

This work is partially supported by NSF 1849238 and Rutgers Global Center. We thank all the anonymous reviewers and the shepherd, Dr. Ilias Leontiadis, for their valuable comments and helpful suggestions.

\end{acks}


\bibliographystyle{ACM-Reference-Format}
\bibliography{ref}

\end{document}